# WAREHOUSING WEB DATA

Jérôme Darmont, Omar Boussaid, and Fadila Bentayeb
Équipe BDD, Laboratoire ERIC
Université Lumière – Lyon 2
5 avenue Pierre Mendès-France
69676 Bron Cedex
France
E-mail: {jdarmont|boussaid|fbentaye}@univ-lyon2.fr

## ABSTRACT

In a data warehousing process, mastering the data preparation phase allows substantial gains in terms of time and performance when performing multidimensional analysis or using data mining algorithms. Furthermore, a data warehouse can require external data. The web is a prevalent data source in this context. In this paper, we propose a modeling process for integrating diverse and heterogeneous (so-called multiform) data into a unified format. Furthermore, the very schema definition provides first-rate metadata in our data warehousing context. At the conceptual level, a complex object is represented in UML. Our logical model is an XML schema that can be described with a DTD or the XML-Schema language. Eventually, we have designed a Java prototype that transforms our multiform input data into XML documents representing our physical model. Then, the XML documents we obtain are mapped into a relational database we view as an ODS (*Operational Data Storage*), whose content will have to be re-modeled in a multidimensional way to allow its storage in a star schema-based warehouse and, later, its analysis.

## 1. INTRODUCTION

The end of the 20th Century has seen the rapid development of new technologies such as web-based, communication, and decision support technologies. Companies face new economical challenges such as e-commerce or mobile commerce, and are drastically changing their information systems design and management methods. They are developing various technologies to manage their data and their knowledge. These technologies constitute what is called "business intelligence". These new means allow them to improve their productivity and to support competitive information monitoring. In this context, the web is now the main farming source for companies, whose challenge is to build web-based information and decision support systems. Our work lies in this field. We present in this paper an approach to build a Decision Support Database (DSDB) whose main data source is the web.

The data warehousing and OLAP (*On-Line Analytical Processing*) technologies are now considered mature in management applications, especially when data are numerical. With the development of the Internet, the availability of various types of data (images, texts, sounds, videos, data from databases…) has increased. These data, which are extremely diverse in nature (we name them "multiform data"), may be unstructured, structured, or even already organized in databases. Their availability in large quantities and their complexity render their structuring and exploitation difficult. Nonetheless, the concepts of data warehousing (Kimball 1996; Inmon 1996; Chaudhuri and Dayal 1997) remain valid for multimedia data (Thuraisingham 2001). In this context, the web may be considered as a farming system providing input to a data warehouse (Hackathorn 2000). Large data volumes and their dating are other arguments in favor of this data webhouse approach (Kimball and Mertz 2000).

Hence, data from the web can be stored into a DSDB such as a data warehouse, in order to be explored by on-line analysis or data mining techniques. However, these multiform data must first be structured into a database, and then integrated in the particular architecture of a data warehouse (fact tables, dimension tables, data marts, data cubes). Yet, the classical warehousing approach is not very adequate when dealing with multiform data. The muldimensional modeling of these data is tricky and may necessitate the introduction of new concepts. Classical OLAP operators may indeed prove inefficient or ill-adapted. Administering warehoused multiform data also requires adequate refreshment strategies when new data pop up, as well as specific physical reorganization policies depending on data usage (to optimize query performance). In order to address these issues, we adopted a step-by-step strategy that helps us handling our problem's complexity.

In a first step, our approach consists in physically integrating multiform data into a relational database playing the role of a buffer ahead of the data warehouse. In a second step, we aim at multidimensionally model these data to prepare them for analysis. The final phase in our process consists in exploring the data with OLAP or data mining techniques.

The aim of this paper is to address the issue of web data integration into a database. This constitutes the first phase in building a multiform data warehouse. We propose a modeling process to achieve this goal. We first designed a conceptual UML model for a complex object representing a superclass of all the types of multiform data we consider (Miniaoui et al. 2001). Note that our objective is not only to store data, but also to truly prepare them for analysis, which is more complex than a mere ETL (*Extracting, Transforming, and Loading*) task. Then, we translated our UML conceptual model into an XML schema definition that represents our logical model. Eventually, this logical model has been instantiated into a physical model that is an XML document. The XML documents we obtain with the help of a Java prototype are mapped into a (MySQL) relational database with a PHP script. We consider this database as an ODS (*Operational Data Storage*), which is a temporary data repository that is typically used in an ETL process before the data warehouse proper is constituted.

The remainder of this paper is organized as follows. Section 2 presents our unified conceptual model for multiform data. Section 3 outlines how this conceptual model is translated into a logical, XML schema definition. Section 4 details how our input data are transformed into an XML document representing our physical model. We finally conclude the paper and discuss future research issues.

## 2. CONCEPTUAL MODEL

The data types we consider (text, multimedia documents,

represents a complex object generalizing all these data types. Note that our goal here is to propose a general data structure: the list of attributes for each class in this diagram is willingly not exhaustive.

A complex object is characterized by its name and its source. The date attribute introduces the notion of successive versions and dating that is crucial in data warehouses. Each complex object is composed of several subdocuments. Each subdocument is identified by its name, its type, its size, and its location (i.e., its physical address). The document type (text, image, etc.) will be helpful later, when selecting an appropriate analysis tool (text mining tools are different from standard data mining tools, for instance). The language class is important for text mining and information retrieval purposes, since it characterizes both documents and keywords.

Eventually, keywords represent a semantic representation of a document. They are metadata describing the object to integrate (medical image, press article...) or its content. Keywords are essential in the indexing process that helps guaranteeing good performances at data retrieval time. Note that we consider only logical indexing here, and not physical relational views from databases) for integration in a data warehouse all bear characteristics that can be used for indexing. The UML class diagram shown in Figure 1 issues arisen by very large amounts of data, which are still quite open as far as we know. Keywords are typically manually captured, but it would be very interesting to mine them automatically with text mining (Tan 1999), image mining (Zhang et al. 2001), or XML mining (Edmonds 2001) techniques, for instance.

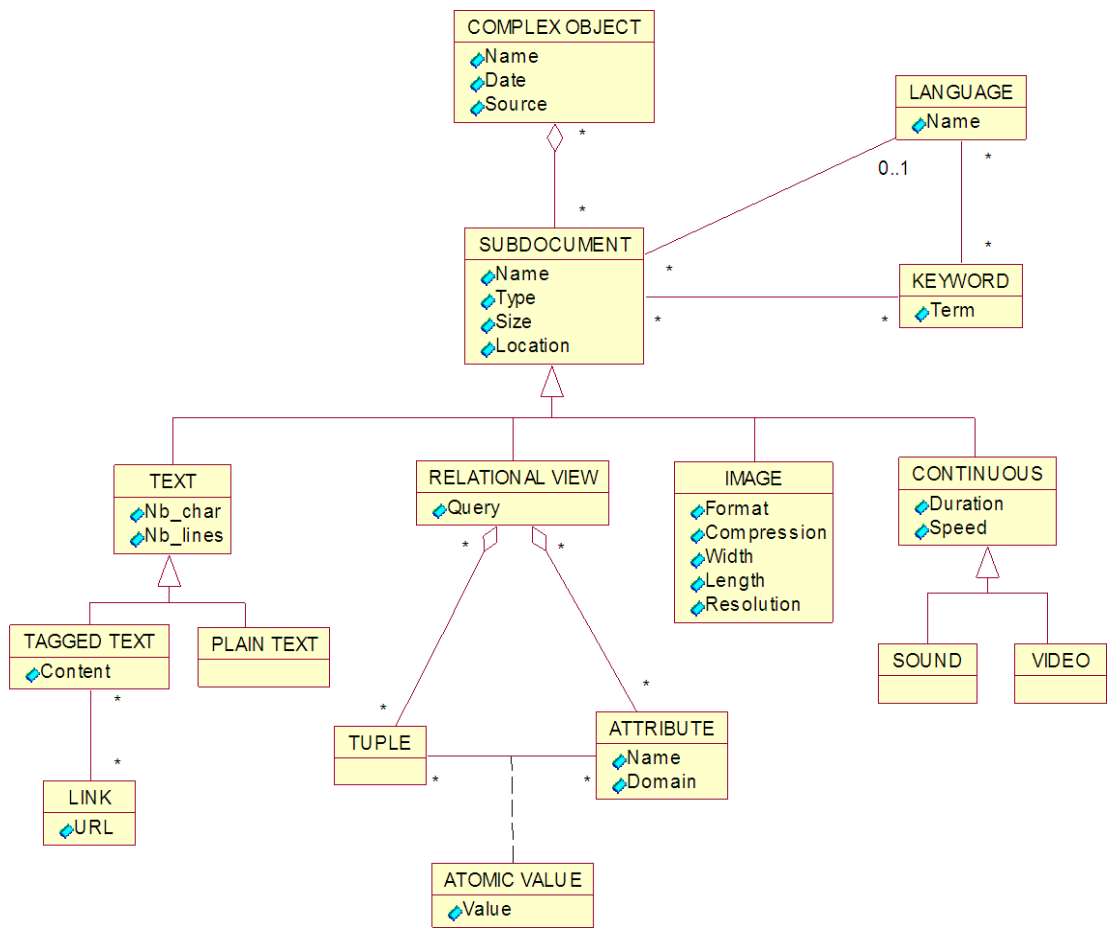

Figure 1: Multiform Data Conceptual Model

All the following classes are subclasses of the subdocument class. They represent the basic data types and/or documents we want to integrate. Text documents are subdivided into plain texts and tagged texts (namely HTML, XML, or SGML documents). Tagged text are further associated to a certain number of links. Since a web page may point to external data (other pages, images, multimedia data, files...), those links help relating these data to their referring page.

Relational views are actually extractions from any type of database (relational, object, object-relational — we suppose a view can be extracted whatever the data model) that will be materialized in the data warehouse. A relational view is a set of attributes (columns, classically characterized by their name and their domain) and a set of tuples (rows). At the intersection of tuples and attributes is a data value. In our model, these values appear as ordinal, but in practice they can be texts or BLOBs containing multimedia data. The query that helped building the view is also stored. Depending on the context, all the data can be stored, only the query and the intention (attribute definitions), or everything. For instance, it might be inadequate to duplicate huge amounts of data, especially if the data source is not regularly updated.

On the other hand, if successive snapshots of an evolving view are needed, data will have to be stored.

Images may bear two types of attributes: some that are usually found in the image file header (format, compression rate, size, resolution), and some that need to be extracted by program, such as color or texture distributions.

Eventually, sounds and video clips are part of a same class because they share continuous attributes that are absent from the other (still) types of data we consider. As far as we know, these types of data are not currently analyzed by mining algorithms, but they do contain knowledge. This is why we take them into account here (though in little detail), anticipating advances in multimedia mining techniques (Thuraisingham 2001).

## 3. LOGICAL MODEL

Our UML model can be directly translated into an XML schema, whether it is expressed as a DTD or in the XML-Schema language. We considered using the XMI method (Cover 2001) to assist us in the translation process, but given the relative simplicity of our models, we proceeded directly. The schema we obtained, expressed as a DTD, is shown in Figure 2.

```
<!ELEMENT COMPLEX_OBJECT (OBJ_NAME, DATE, SOURCE, SUBDOCUMENT+)>
  <!ELEMENT OBJ_NAME (#PCDATA)>
  <!ELEMENT DATE (#PCDATA)>
  <!ELEMENT SOURCE (#PCDATA)>
  <!ELEMENT SUBDOCUMENT (DOC_NAME, TYPE, SIZE, LOCATION, LANGUAGE?,
KEYWORD*, (TEXT | RELATIONAL_VIEW | IMAGE | CONTINUOUS))>
    <!ELEMENT DOC_NAME (#PCDATA)>
    <!ELEMENT TYPE (#PCDATA)>
    <!ELEMENT SIZE (#PCDATA)>
    <!ELEMENT LOCATION (#PCDATA)>
    <!ELEMENT LANGUAGE (#PCDATA)>
    <!ELEMENT KEYWORD (#PCDATA)>
    <!ELEMENT TEXT (NB_CHAR, NB_LINES, (PLAIN_TEXT | TAGGED_TEXT))>
        <!ELEMENT NB_CHAR (#PCDATA)>
        <!ELEMENT NB_LINES (#PCDATA)>
        <!ELEMENT PLAIN_TEXT (#PCDATA)>
        <!ELEMENT TAGGED_TEXT (CONTENT, LINK*)>
            <!ELEMENT CONTENT (#PCDATA)>
            <!ELEMENT LINK (#PCDATA)>
    <!ELEMENT RELATIONAL_VIEW (QUERY?, ATTRIBUTE+, TUPLE*)>
        <!ELEMENT QUERY (#PCDATA)>
        <!ELEMENT ATTRIBUTE (ATT_NAME, DOMAIN)>
            <!ELEMENT ATT_NAME (#PCDATA)>
            <!ELEMENT DOMAIN (#PCDATA)>
        <!ELEMENT TUPLE (ATT_NAME_REF, VALUE)+>
            <!ELEMENT ATT_NAME_REF (#PCDATA)>
            <!ELEMENT VALUE (#PCDATA)>
    <!ELEMENT IMAGE (COMPRESSION, FORMAT, RESOLUTION, LENGTH, WIDTH)>
        <!ELEMENT FORMAT (#PCDATA)>
        <!ELEMENT COMPRESSION (#PCDATA)>
        <!ELEMENT LENGTH (#PCDATA)>
        <!ELEMENT WIDTH (#PCDATA)>
        <!ELEMENT RESOLUTION (#PCDATA)>
    <!ELEMENT CONTINUOUS (DURATION, SPEED, (SOUND | VIDEO))>
        <!ELEMENT DURATION (#PCDATA)>
        <!ELEMENT SPEED (#PCDATA)>
        <!ELEMENT SOUND (#PCDATA)>
        <!ELEMENT VIDEO (#PCDATA)>
```

Figure 2: Logical Model (DTD)

We applied minor shortcuts not to overload this schema. Since the language, keyword, link and value classes only bear one attribute each, we mapped them to single XML elements, rather than having them be composed of another, single element. For instance, the language class became the language element, but this element is not further composed of the name element. Eventually, since the attribute and the tuple elements share the same sub-element "attribute name", we labeled it ATT_NAME in the attribute element and ATT_NAME_REF (reference to an attribute name) in the tuple element to avoid any confusion or processing problem.

## 4. PHYSICAL MODEL

We have developed a Java prototype capable of taking as input a data source from the web, fitting it in our model, and producing an XML document. The source code of this application we baptized web2xml is available on-line: http://bdd.univ-lyon2.fr/download/web2xml.zip. We view the XML documents we generate as the final physical models in our process.

The first step of our multiform data integration approach consists in extracting the attributes of the complex object that has been selected by the user. A particular treatment is applied depending on the subdocument class (image, sound, etc.), since each subdocument class bears different attributes. We used three ways to extract the actual data: (1) manual capture by the user, through graphical interfaces; (2) use of standard Java methods and packages; (3) use of ad-hoc automatic extraction algorithms. Our objective is to progressively reduce the number of manually-captured attributes and to add new attributes that would be useful for later analysis and that could be obtained with data mining techniques.

The second step when producing our physical model consists in generating an XML document. The algorithm's principle is to parse the schema introduced in Figure 2 recursively, fetching the elements it describes, and to write them into the output XML document, along with the associated values extracted from the original data, on the fly. Missing values are currently treated by inserting an empty element, but strategies could be devised to solve this problem, either by prompting the user or automatically.

At this point, our prototype is able to process all the data classes we identified in Figure 1. Figure 3 illustrates how one single document (namely, an image) is transformed using our approach. A composite document (such as a web page including pieces of text, XML data, data from a relational database, and an audio file) would bear the same form. It would just have several different subdocument elements instead of one (namely plain and tagged text, relational view, and continuous/sound subdocuments, in our example).

Eventually, in order to map XML documents into a (MySQL) relational database, we designed a prototype baptized xml2rdb. This PHP script is also available on-line: http://bdd.univ-lyon2.fr/xml2rdb/. It operates in two steps. First, a DTD parser exploits our logical model (Figure 2) to build a relational schema, i.e., a set of tables in which any valid XML document (regarding our DTD) can be mapped. To achieve this goal, we mainly used the techniques proposed by (Anderson et al. 2000; Kappel at al. 2000). Note that our DTD parser is a generic tool: it can operate on any DTD. It takes into account all the XML element types we need, e.g., elements with +, *, or ? multiplicity, element lists, selections, etc. The last and easiest step consists in loading a valid XML document into the previously build relational structure.

## 5. CONCLUSION AND PERSPECTIVES

We presented in this paper a modeling process for integrating multiform data from the web into a Decision Support

| Image | XML Model |
|---|---|
| 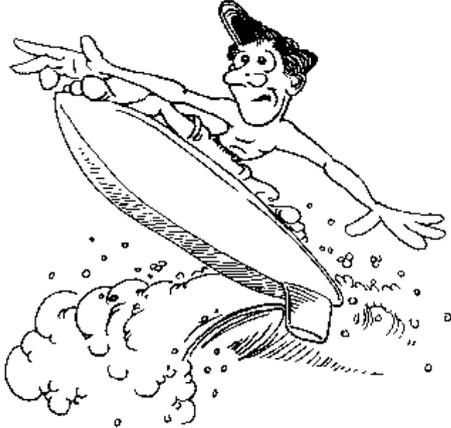 User-prompted keywords:<br>— surf<br>— black and white<br>— wave | ```<br><?XML version=1.0?><br><!DOCTYPE MlfDt SYSTEM "mlfd.dtd"><br><COMPLEX_OBJECT><br>    <OBJ_NAME>Sample image</OBJ_NAME><br>    <DATE>2002-06-15</DATE><br>    <SOURCE>Local</SOURCE><br>    <SUBDOCUMENT><br>        <DOC_NAME>Surf</DOC_NAME><br>        <TYPE>Image</TYPE><br>        <SIZE>4407</SIZE><br>        <LOCATION>gewis_surfer2.gif</LOCATION><br>        <KEYWORD>surf</KEYWORD><br>        <KEYWORD>black and white</KEYWORD><br>        <KEYWORD>wave</KEYWORD><br>        <IMAGE><br>            <FORMAT>Gif</FORMAT><br>            <COMPRESSION/><br>            <WIDTH>344</WIDTH><br>            <LENGTH>219</LENGTH><br>            <RESA>72dpi</RESA><br>        </IMAGE><br>    </SUBDOCUMENT><br></COMPLEX_OBJECT><br>``` |

Figure 3: Sample Physical Model for an Image

Database such as a data warehouse. Our conceptual UML model represents a complex object that generalizes the different multiform data that can be found on the web and that are interesting to integrate in a data warehouse as external data sources. Our model allows the unification of these different data into a single framework, for purposes of storage and preparation for analysis. Data must indeed be properly "formatted" before OLAP or data mining techniques can apply to them.

Our UML conceptual model is then directly translated into an XML schema (DTD or XML-Schema), which we view as a logical model. The last step in our (classical) modeling process is the production of a physical model in the form of an XML document. XML is the format of choice for both storing and describing the data. The schema indeed represents the metadata. XML is also very interesting because of its flexibility and extensibility, while allowing straight mapping into a more conventional database if strong structuring and retrieval efficiency are needed for analysis purposes.

The aim of the first step in our approach was to structure multiform data and integrate them in a database. At this point, data are managed in a transactional way: it is the first modeling step. Since the final objective of our approach is to analyze multimedia data, and more generally multiform data, it is necessary to add up a layer of multidimensional modeling to allow an easy and efficient analysis.

One first improvement on our work could be the use of the XML-Schema language instead of a DTD to describe our logical model. We could indeed take advantage of XML-Schema's greater richness, chiefly at the data type diversity and (more importantly) inheritance levels. This would facilitate the transition between the UML data representation and the XML data representation.

Both the XML and XML-Schema formalisms could also help us in the multidimensional modeling of multiform data, which constitutes the second level of structuring. In opposition to the classical approach, designing a multidimensional model of multiform data is not easy at all. A couple of studies deal with the multidimensional representation and have demonstrated the feasibility of UML snowflake diagrams (Jensen et al. 2001), but they remain few.

We are currently working on the determination of facts in terms of measures and dimensions with the help of imaging, statistical, and data mining techniques. These techniques are not only useful as analysis support tools, but also as modeling support tools. Note that it is not easy to refer to measures and dimensions when dealing with multimedia documents, for instance, without having some previous knowledge about their content. Extracting semantics from multimedia documents is currently an open issue that is addressed by numerous research studies. We work on these semantic attributes to build multidimensional models.

The diversity of multiform data also requires adapted operators. Indeed, how can we aggregate data that are not quantitative, such as most multimedia attributes? Classical OLAP operators are not adapted to this purpose. We envisage to integrate some data mining tasks (e.g., clustering) as new OLAP aggregation operators. Multiform data analysis with data mining techniques can also be complemented with OLAP's data navigation operators.

Eventually, using data mining techniques help us addressing certain aspects of data warehouse auto-administration. They indeed allow to design refreshment strategies when new data pop up and to physically reorganize the data depending on their usage in order to optimize query performance. This aspect is yet another axis in our work that is necessary in our global approach.